\documentclass[12pt]{article}

\usepackage{amsmath}
\begin{document}

\title{Nonlinear Shear-free Radiative Collapse}
\author{S.S. Misthry$^
{\footnote{SSM and SDM dedicate this work to Peter Leach on his
65$^{\text{th}}$ birthday}}$ $^{\footnote{Permanent address:
Department of Mathematics, Durban University of Technology,
Steve Biko Campus, Durban, 4001}}$, S.D. Maharaj$^*$ and P.G.L. Leach, \\
Astrophysics and Cosmology Research Unit,\\
School of Mathematical Sciences, University of KwaZulu-Natal,\\ Private Bag X54001, Durban 4000, South Africa}

\maketitle

\begin{abstract}
We study realistic models of relativistic radiating
stars undergoing gravitational collapse which have
vanishing Weyl tensor components. Previous investigations
are generalised by retaining the inherent nonlinearity at
the boundary. We transform the boundary condition to
an Abel equation of the first kind. A variety of nonlinear
solutions are generated all of which can be written explicitly.
Several classes of infinite solutions exist.
\\
\\
{\it{Keywords}}: astrophysics; Einstein equations; radiating stars.
\\ \\
AMS numbers: 83 C 15; 83 C 22; 85 A 99
\end{abstract}

\newcommand{\be}{\begin{equation}}
\newcommand{\ee}{\end{equation}}
\newcommand{\bea}{\begin{eqnarray}}
\newcommand{\eea}{\end{eqnarray}}
\newcommand{\et}{\it{et al.}}
\newcommand{\bib}{\bibitem}

\section{Introduction}
The evolution of a radiating star undergoing gravitational collapse,
in the context of general relativity, has occupied the attention of
researchers in astrophysics in recent times. The derivation of the
junction conditions by Santos \cite{santos} has made it possible to
obtain exact models of an interior spacetime with heat flux to
match with the exterior Vaidya spacetime; at the boundary of the
star the radial pressure is nonzero. A variety of exact solutions
has been generated over the years to study the cosmic censorship
hypothesis, gravitational collapse with dissipation, end state of
superdense matter, dynamical stability of radiating matter and
temperature profiles in the context of irreversible thermodynamics.
De Oliviera {\it{et al.}} \cite{deOliviera} proposed a radiating
model in which an initial static configuration leads to collapse.
This approach may be adapted to study the end state of collapse as
shown by Govender {\it{et al.}} \cite{govender1}. Kolassis {\it{et
al.}} \cite{kolassis} assumed that the fluid trajectories are
geodesic and generated exact solutions. These assumptions lead to
particular solutions which may be used to study the physical
features of the model such as the relaxational effects of the
collapsing fluid on the temperature profile in theories of causal
thermodynamics \cite{DiPrisco1}-\cite{govender4}.

In a recent treatment Herrera {\et} \cite{herrera2} proposed a model
in which the form of the Weyl tensor was highlighted when studying
radiative collapse. This approach has the advantage of simplifying
the Einstein field equations. However, Herrera {\et} were not able to solve
the junction conditions; only an approximate solution was found.
Maharaj and Govender \cite{maharaj} showed that it is possible to
solve the field equations and the junction conditions exactly. Their
solution is expressible in terms of elementary functions and
contains the Friedmann dust solution as a special case. It is
interesting to note that Herrera {\et} \cite{herrera3} showed that
other classes of solutions in terms of the elementary functions are
possible. The exact solutions in both \cite{maharaj} and
 \cite{herrera3} depend upon the introduction of a transformation that
linearises the boundary condition. The purpose of this paper is to
demonstrate that it is possible to obtain other models by
transforming the boundary condition to an Abel's equation which is
necessarily nonlinear. We explicitly find exact solutions to the
Abel equation under particular assumptions and thereby demonstrate
that conformally flat radiating stars contain a richer structure
than previously suspected.

The main objective of this paper is to show that we can generate
radiating relativistic stellar models without having to eliminate
the nonlinearity at the boundary. In Section 2, we describe the
basic features of the model for a radiating star and present the
relevant differential equations. Results generated in previous
investigations are briefly discussed in Section 3. These have been
obtained by introducing  a transformation that leads to a linear
equation at the boundary. In Section 4, we introduce a new
transformation at the boundary that leads to an Abel's equation. We
show explicitly that a variety of exact solutions can be generated
from the Abel equation. Consequently a variety of new models for
radiating relativistic stars, with vanishing Weyl stresses, are
possible. The physical features of the solutions are briefly
considered in Section 5.
\section{The Model}
We consider a spherically symmetric radiating star undergoing
shear-free gravitational collapse. The line element for shear-free
matter interior to the boundary of the radiating star is given by
\begin{equation}
ds^2=-A^2dt^2+B^2[dr^2+r^2(d\theta^2+\sin^2\theta d\phi^2)]\label{line1}
\end{equation}
where $A=A(t,r)$ and $B=B(t,r)$ are the metric functions. The energy momentum tensor including radiation for the interior spacetime is\\
\be
T_{ab}=(\rho+p)u_au_b+pg_{ab}+q_au_b+q_bu_a \label{E-P}
\ee
where the energy density $\rho$, the pressure $p$ and the heat flow vector $q_a$ are measured relative to the timelike fluid 4-velocity $u^a=\frac{1}{A}\delta^{a}_{0}.$ The heat flow vector assumes the form $q^a=(0,q,0,0)$ since $q^a u_a=0$ for radially directed heat flow.

The nonzero components of the Einstein field equations, for the
line element $(\ref{line1})$ and the energy momentum (\ref{E-P}),
can be written as
\begin{subequations} \label{EFI1}
\bea
\rho&=&\frac{3}{A^2}\frac{\dot{B}^2}{B^2}-\frac{1}{B^2}\left(2\frac{B''}{B}-\frac{B'^2}{B^2}+\frac{4}{r}\frac{B'}{B}\right)\label{EFI1a}\\
p&=&\frac{1}{A^2}\left(-2\frac{\ddot{B}}{B}-\frac{\dot{B}^2}{B^2}+2\frac{\dot{A}}{A}\frac{\dot{B}}{B}\right) \nonumber \\
& &
+\frac{1}{B^2}\left(\frac{B'^2}{B^2}+2\frac{A'}{A}\frac{B'}{B}+\frac{2}{r}\frac{A'}{A}+\frac{2}{r}\frac{B'}{B}\right)\label{EFI1b}\\
p&=&-2\frac{1}{A^2}\frac{\ddot{B}}{B}+2\frac{\dot{A}}{A^3}\frac{\dot{B}}{B}-\frac{1}{A^2}\frac{\dot{B}^2}{B^2}+\frac{1}{r}\frac{A'}{A}\frac{1}{B^2}\nonumber\\
& &
+\frac{1}{r}\frac{B'}{B^3}+\frac{A''}{A}\frac{1}{B^2}-\frac{B'^2}{B^4}+\frac{B''}{B^3} \label{EFI1c} \\
q&=&-\frac{2}{AB^2}\left(-\frac{\dot{B'}}{B}+\frac{B'\dot{B}}{B^2}+\frac{A'}{A}\frac{\dot{B}}{B}\right)\label{EFI1d}
\eea
\end{subequations}
\\
The Weyl tensor has all components proportional to \be
C_{2323}=\frac{r^4}{3}B^2\sin^2\theta
\left[\left(\frac{A'}{A}-\frac{B'}{B}\right)\left(\frac{1}{r}+2\frac{B'}{B}\right)-\left(\frac{A''}{A}-\frac{B''}{B}\right)\right]\nonumber
\ee according to
\bea
C_{2323}&=&-r^4\left(\frac{B}{A}\right)^2\sin^2\theta C_{0101}=2r^2\left(\frac{B}{A}\right)^2 \sin^2 \theta C_{0202} \nonumber \\
&=&2r^2\left(\frac{B}{A}\right)^2C_{0303}=-2r^2\sin^2 \theta
C_{1212}=-2r^2C_{1313}\nonumber \eea
\\
which represent the tidal forces. For conformal flatness these
components must all vanish so that $C_{2323}=0$. This leads to a
nonlinear partial differential equation which is easily solved so
that
\be
A=(C_1(t)r^2+1)B. \label{A}
\ee
Now from (\ref{EFI1b}) and (\ref{EFI1c}), and using (\ref{A}), we obtain
\be
\frac{B''}{B'}-2\frac{B'}{B}-\frac{1}{r}=0 \label{pr-isot}
\ee which
is the condition of pressure isotropy. Equation (\ref{pr-isot}) is
integrable and we get
\be B=\frac{1}{C_2(t)r^2+C_3(t)}\label{B}
\ee
where $C_1(t), C_2(t)$ and $C_3(t)$ are functions of time. The forms
for the metric functions $A$ and $B$ given above generate an exact
solution to the Einstein field equations (\ref{EFI1}).

The interior spacetime $(\ref{line1})$ has to be matched across the
boundary $r=b$ to the exterior Vaidya spacetime
\be
ds^2=-\left(1-\frac{2m(v)}{R}\right)dv^2-2dv dR +
R^2(d\theta^2+\sin^2\theta d\phi^2) \label{vaidya}
\ee
The hypersurface at the boundary is denoted by $\Sigma$. The junction
conditions at $\Sigma$ have the form

\begin{subequations}\label{junct}
\bea (Adt)_{\Sigma} &=&
\left[\left(1-\frac{m}{R}+2\frac{dR}{dv}\right)^{1/2}dv\right]_{\Sigma} \label{juncta}\\
(rB)_{\Sigma}&=&R_{\Sigma} \label{junctb}\\
p_{\Sigma}&=&(qB)_{\Sigma}\label{junctc}\\
\left[m(v)\right]_{\Sigma}&=&
\left[\frac{r^3}{2}\left(\frac{\dot{B}^{2}B}{A^2}-\frac{B'^2}{B}\right)-r^2B'\right]_{\sum}
\label{junctd} \eea
\end{subequations}
on matching (\ref{line1}) and (\ref{vaidya}). For our model the
junction conditions (\ref{junct}) reduce to the following nonlinear
ordinary differential equation
\begin{eqnarray}
\ddot{C_2}b^2+\ddot{C_3}-\frac{3}{2}\frac{(\dot{C_2}b^2+\dot{C_3})^2}{C_2b^2+C_3}-\frac{\dot{C_1}b^2(\dot{C_2}b^2+\dot{C_3})}{C_1 b^2+1}-2(\dot{C_3}C_1-\dot{C_2})b\nonumber\\
+2\frac{C_1b^2+1}{C_2b^2+C_3}\left[C_2(C_2-2C_1C_3)b^2+C_3(C_1C_3-2C_2)\right]=0 \label{geqn}
\end{eqnarray}
\\
resulting from the (nonvanishing) pressure gradient across the
hypersurface $\Sigma$. Equation (\ref{geqn}) governs the evolution
of a radiating star with vanishing Weyl stresses. To complete the
description of this radiating model we need to solve the remaining
junction condition (\ref{geqn}).
\section {Elementary solutions}
The governing equation (\ref{geqn}) is a highly nonlinear equation
presenting a formidable mathematical task to solve exactly in
general. Note that in previous attempts to integrate (\ref{geqn})
assumptions were made that effectively linearised this boundary
condition. We briefly summarise the known results.

Herrera {\it{et al.}} \cite{herrera2} assumed the following
approximate forms for the temporal functions \be C_1=\epsilon
c_1(t),\,\,\,\,\, C_2=0,\,\,\,\,\, C_3=\frac{a}{t^2},\label{assum1}
\ee where $ 0 < \epsilon << 1$ and $a > 0$, is a constant.  With the
assumptions contained in (\ref{assum1}), (\ref{geqn}) yields the
approximate solution \be
 C_1\approx C_1(0)\exp
\left({\frac{-t^2}{2b^2}-\frac{2t}{b}}\right)\nonumber
\ee
Note that on setting $C_1 = 0$ the solution reduces to a collapsing
Friedmann dust sphere.

Maharaj and Govender \cite{maharaj} were the first to determine a closed form solution for (\ref{geqn}). They assumed that $C_1 = C$ (a constant), $C_2
= 0$ and introduced the transformation $C_3\equiv u^{-2}$ so that
(\ref{geqn}) takes the linear form \be
\ddot{u}-2Cb\dot{u}-(Cb^2+1)Cu=0 \nonumber \ee
in the new variable {\it{u}}. Three categories of closed form solutions in
terms of the elementary functions were obtained depending on the
nature of the roots of the characteristic equation.

Herrera {\it{et al.}} \cite{herrera3} extended this treatment to
obtain a wider class of solutions. They set $C_1 = C$ (a constant),
$C_2=\alpha C_3$ and introduced the transformation
$C_3(t)=u^{-2}(t)$ so that (\ref{geqn}) can be written as
\\
\be \ddot{u}-\frac{2(C-\alpha)b}{\alpha
b^2+1}\dot{u}-\frac{(Cb^2+1)}{(\alpha
b^2+1)^2}[\alpha(\alpha-2C)b^2 +(C-2\alpha)]u=0 \label{herrerasol}
\ee which is linear in $u$. Then equation (\ref{herrerasol}) admits
three classes of solution given by\\ \\
{\it{Case 1}}:\,\,$(C-\alpha)^2b^2+(Cb^2+1)[\alpha(\alpha-2C)b^2+(C-2\alpha)]>0$\\
\bea
C_3(t)=\left[\beta_1\exp \left(\frac{(C-\alpha)b
+\sqrt{(C-\alpha)^2b^2+(Cb^2+1)[\alpha(\alpha-2C)b^2
+(C-2\alpha)]}}{\alpha b^2+1}\right)t\ \right.\nonumber \\
\left.+\beta_2\exp{\left(\frac{(C-\alpha)b
-\sqrt{(C-\alpha)^2b^2+(Cb^2+1)[\alpha(\alpha-2C)b^2
+(C-2\alpha)]}}{\alpha b^2+1}\right)t}\right]^{-2}  \nonumber \eea\\
{\it{Case 2}}:\,\,$(C-\alpha)^2b^2+(Cb^2+1)[\alpha(\alpha-2C)b^2+(C-2\alpha)]<0$\\
\bea C_3(t)=\left[\exp\left({\frac{(C-\alpha)b}{\alpha
b^2+1}t}\right)\left(\beta_1\cos\left({\frac{\sqrt{(C-\alpha)^2b^2+(Cb^2+1)[\alpha(\alpha-2C)b^2
+(C-2\alpha)]}}{(\alpha b^2+1)}}\right)t \right.\right.\nonumber \\
\left.\left.\left.+\beta_2\sin\left({\frac{\sqrt{(C-\alpha)^2b^2+(Cb^2+1)[\alpha(\alpha-2C)b^2
+(C-2\alpha)]}}{(\alpha
b^2+1)}}\right.\right)t\right)\right]^{-2}\nonumber \eea\\
{\it{Case 3}}:\,\,$(C-\alpha)^2b^2+(Cb^2+1)[\alpha(\alpha-2C)b^2+(C-2\alpha)]=0$\\
\bea C_3(t)=(\beta_1+\beta_2t)^{-2}\exp\left({\frac{-2(C-\alpha)b}{\alpha
b^2+1}t}\right)\nonumber \eea\\ where $ \beta_1 $ and $ \beta_2$ are constants of
integration. These solutions reduce to the  Maharaj and Govender
\cite{maharaj} model when $\alpha = 0$. Note that other
transformations that linearise (\ref{geqn}) are possible as indicated
in \cite{herrera3}.
\section{Abel Equation}
The nonlinearity and complexity in the boundary condition (\ref{geqn})
is clearly evident. It is therefore remarkable that closed form solutions
in terms of elementary functions have been shown to exist as
shown in Section 3. These particular closed form solutions have been
generated from linearised forms of the governing equation
(\ref{geqn}). A natural extension would be a study of the existence
of nonlinear solutions to the differential equation (\ref{geqn}). Such
classes of solutions, if they exist, are important in the study of
nonlinear behaviour of the shear-free, conformally flat model.
Consequently we seek classes of solutions which retain the inherent
nonlinear structure of (\ref{geqn}). These have not been found in
the past due to the inherent difficulties of coping with nonlinearity.

Here we consider a particular nonlinear transformation which leads
to exact solutions. It is convenient to replace the function
$C_1(t)$ with \be U = C_1b^2+1\label{Atrans} \ee Then the governing
equation (\ref{geqn}) may be written with some rearrangement as
\begin{eqnarray}
\dot{U}(\dot{C_2}b^2+\dot{C_3}) + U\left[\frac{3}{2}\frac{(\dot{C_2}b^2+\dot{C_3})^2}{C_2b^2+C_3}-\frac{2}{b}(\dot{C_2}b^2+\dot{C_3})-(\ddot{C_2}b^2+\ddot{C_3})\right]\nonumber \\+2U^2\left[\frac{\dot{C_3}}{b}-\frac{1}{C_2b^2+C_3}(C_2^2b^2-\frac{C_3^2}{b^2})\right]+2U^3\frac{2C_2b^2-C_3}{C_2b^2+C_3}.\frac{C_3}{b^2}=0 \label{Aeqn}
\end{eqnarray}
Equation (\ref{Aeqn}) is complicated, but has the generic structure
\\
\be {\cal A}\dot{U}+{\cal B}U+{\cal C}U^2+{\cal D}U^3=0
\label{calAeqn} \ee where we have set \bea
{\cal A}&=&\dot{C_2}b^2+\dot{C_3}\nonumber\\
{\cal B}&=&\frac{3}{2}\frac{(\dot{C_2}b^2+\dot{C_3})^2}{C_2b^2+C_3}-\frac{2}{b}(\dot{C_2}b^2+\dot{C_3})-(\ddot{C_2}b^2+\ddot{C_3})\nonumber\\
{\cal C}&=&2\left(\frac{\dot{C_3}}{b}-\frac{1}{C_2b^2+C_3}(C_2^2b^2-\frac{C_3^2}{b^2})\right)\nonumber\\
{\cal
D}&=&2\left(\frac{2C_2b^2-C_3}{C_2b^2+C_3}.\frac{C_3}{b^2}\right)\nonumber
\eea \\The transformed equation (\ref{calAeqn}) is an Abel's equation
of the first kind in the variable $U$. Abelian equations are
difficult to solve in general. However, the advantage of utilising
the transformation (\ref{Atrans}) is that (\ref{calAeqn}) is a first
order differential equation in $U$. In the following we present a
comprehensive mathematical treatment of (\ref{calAeqn}) and derive
several classes of solutions.
\subsection{Case 1: ${ \cal A}=0$}
The restriction ${\cal A}=0$ immediately gives \be C_2b^2+C_3=\alpha
\label{cons1} \ee where $\alpha$ is a constant of integration. Then
$(\ref{Aeqn})$ becomes \be
2U^2\left[\frac{\dot{C_3}}{b}-\frac{1}{\alpha}\left(C_2^2b^2-\frac{C_3^2}{b^2}\right)\right]+2U^3\frac{2C_2b^2-C_3}{\alpha}.\frac{C_3}{b^2}=0
\label{algeqn}
\ee which is an algebraic equation in $U$.
\\

Two cases arise: $U=0$ or $U\neq0$ in (\ref{algeqn}). We easily
find:
\begin{subequations}\label{solcase1}
\bea
C_1&=&\left\{
\begin{array}{lll}
-\frac{1}{b^2}& &,\,\,\, U=0\\
\frac{\alpha}{C_3(2\alpha-3C_3)}\left(\frac{\alpha}{b^2}-\frac{4C_3}{b^2}+\frac{3C_3^2}{\alpha
b^2}-\frac{\dot{C_3}}{b}\right)& &,\,\,\, U\neq0\\
\end{array}\right.\\
C_2&=&\frac{\alpha-C_3}{b^2} \\
C_3&=& \text{arbitrary function of time}
\eea
\end{subequations}
\\
This solution is particularly attractive since we have an infinite choice of $C_3$ and no
integration is required.
\\
\subsection{Case 2: ${ \cal D}=0$}
With ${\cal{D}}=0$ we have two possibilities: either $2C_2b^2-C_3=0$
or $C_3=0$.\\

We firstly consider $2C_2b^2-C_3=0$. Then (\ref{Aeqn}) becomes \be
\dot{U}+U\left[\frac{3}{2}\frac{\dot{C_3}}{C_3}-\frac{2}{b}-\frac{\ddot{C_3}}{\dot{C_3}}\right]=-U^2\left[\frac{4}{3b}+\frac{2}{3}\frac{C_3}{b^2\dot{C_3}}\right]\nonumber
\ee This is a Bernoulli equation with solution \be
U=\frac{\dot{C_3}C_3^{-3/2}e^{2t/b}}{K-\frac{8}{3b}e^{2t/b}C_3^{-1/2}+\frac{6}{b^2}\int
C_3^{-1/2}e^{2t/b}dt}\label{ucase2a}\nonumber \ee where $K$ is a
constant of
integration. Hence for this first case we have the solution\\
\begin{subequations}\label{solcase2a}
\bea
C_1&=&\frac{1}{b^2}\left(\frac{\dot{C_3}C_3^{-3/2}e^{2t/b}}{K-\frac{8}{3b}e^{2t/b}C_3^{-1/2}+\frac{6}{b^2}\int C_3^{-1/2}e^{2t/b}dt}-1\right)\\
C_2&=&\frac{C_3}{2b^2} \\
C_3&=& \text{arbitrary function of time} \eea
\end{subequations}
\\
This is an infinite class of solutions depending on $C_3$.

Now we consider $C_3=0$. The Abel equation $(\ref{Aeqn})$ becomes
\be \dot{U}+U\left(\frac{3}{2}\frac{\dot{C_2}}{C_2}
-\frac{2}{b}-\frac{\ddot{C_2}}{\dot{C_2}}\right)=2U^2\frac{C_2}{\dot{C_2}b^2}\nonumber
\ee This is again a Bernoulli equation with solution \be
U=\frac{\dot{C_2}C_2^{-3/2}e^{2t/b}}{K'-\frac{2}{b^2}\int
e^{2t/b}C_2^{-1/2}dt}\label{ucase2b}\nonumber \ee where $K'$ is a
constant of integration. Therefore for the second case we have the
solution
\begin{subequations}\label{solcase2b}
\bea
C_1&=&\frac{1}{b^2}\left(\frac{\dot{C_2}C_2^{-3/2}e^{2t/b}}{K'-\frac{2}{b^2}\int
e^{2t/b}C_2^{-1/2}dt}-1\right)\\ C_2&=& \text{arbitrary function of
time}\\
C_3&=&0
 \eea
\end{subequations}
\\
Again we have generated an infinite class of solutions depending on
$C_2$.
\subsection{Case 3: ${ \cal C}=0$}
Upon setting ${\cal{C}}=0$ we obtain the equation \be
\frac{\dot{C_3}}{b}-\frac{1}{C_2b^2+C_3}\left(C_2^2b^2-\frac{C_3^2}{b^2}\right)=0\nonumber
\ee
\\
This equation is quadratic in $C_2$ which implies \be
C_2=\frac{\dot{C_3}b\pm\sqrt{\dot{C_3}^2b^2-4C_3(C_3+\dot{C_3}b)}}{2b^2}\label{cons2}\nonumber
\ee
\\
Hence $C_2$ is a known quantity if the function $C_3$ is specified.

The Abelian equation (\ref{Aeqn}) has the form
\bea &
&\dot{U}(\dot{C_2}b^2+\dot{C_3})
+U\left[\frac{3}{2}\frac{(\dot{C_2}b^2+\dot{C_3})^2}{C_2b^2+C_3}-\frac{2}{b}(\dot{C_2}b^2+\dot{C_3})\right.\nonumber\\
& & \left.-(\ddot{C_2}b^2+\ddot{C_3})\right]
=-2U^3\left[\frac{2C_2b^2-C_3}{C_2b^2+C_3}.\frac{C_3}{b^2}\right]\nonumber
\eea The equation is complicated, but may be written concisely as \be
\alpha \dot{U}+\beta U = -\gamma U^3 \label{case3}\ee
where
\bea
\alpha &=& \dot{C_2}b^2+\dot{C_3}\nonumber\\
\beta &=& \frac{3}{2}\frac{(\dot{C_2}b^2+\dot{C_3})^2}{C_2b^2+C_3}-\frac{2}{b}(\dot{C_2}b^2+\dot{C_3})
-(\ddot{C_2}b^2+\ddot{C_3})\nonumber\\
\gamma &=&2\frac{2C_2b^2-C_3}{C_2b^2+C_3}.\frac{C_3}{b^2}\nonumber
\eea The simpler equation (\ref{case3}) has the form of a Bernoulli
equation with solution \bea U&=&\frac{1}{e^{\int (\beta/\alpha)
dt}\left(\int \frac{2\gamma}{\alpha}e^{-\int
(2\beta/\alpha)dt}dt\right)^{1/2}}\nonumber \\
&=&\frac{e^{(2t/b)}(\dot{C_2}b^2+\dot{C_3})}{(C_2b^2+C_3)^{3/2}
\left[K^{''}+\frac{4}{b^2}\int {\frac{e^{(4t/b)}C_3(2C_2b^2-C_3)
(\dot{C_2}b^2+\dot{C_3})}{(C_2b^2+C_3)^4}}dt\right]^{1/2}} \eea
\\
where $K^{''}$ is a constant of integration. Consequently for this
case we have the solution
\begin{subequations}\label{solcase3}
\bea
C_1&=&\frac{1}{b^2}\left(\frac{e^{2t/b}(\dot{C_2}b^2+\dot{C_3})}{(C_2b^2+C_3)^{3/2}
\left[K^{''}+\frac{4}{b^2}\int {\frac{e^{(4t/b)}C_3(2C_2b^2-C_3)
(\dot{C_2}b^2+\dot{C_3})}{(C_2b^2+C_3)^4}}dt\right]^{1/2}}-1\right)\\
C_2&=&\frac{\dot{C_3}b\pm\sqrt{\dot{C_3}^2b^2-4C_3(C_3+\dot{C_3}b)}}{2b^2}\\
C_3&=&\text{arbitrary function of time} \eea
\end{subequations}\\
Again an infinite class of solutions is possible.
\subsection{Case 4: }
This is the most general case and corresponds to the situation for which
all of the coefficients ${\cal A}, {\cal B}, {\cal C}$ and ${\cal D}$
are nonzero. Equation $(\ref{calAeqn})$ can be written
as

\be \dot{U}=-\frac{\cal B}{\cal A}U-\frac{\cal C}{\cal
A}U^2-\frac{\cal D}{\cal A}U^3 \label{quadeqn}\nonumber \ee
\\
so that a variables separable equation is possible if $ \frac{\cal
B}{\cal A}, \frac{\cal C}{\cal A}$ and $ \frac{\cal D}{\cal A}$ are
constants. Then the solution may be written as the quadrature
\be
t-t_0=\int{\frac{dU}{\frac{\cal B}{\cal A}U+\frac{\cal C}{\cal
A}U^2+\frac{\cal D}{\cal A}U^3}} \label{quadeqn}\ee
\\
It is important to emphasize that the additional constraints
generated by $ \frac{\cal B}{\cal A}, \frac{\cal C}{\cal A}$ and $
\frac{\cal D}{\cal A}$ being constant simultaneously are not easy to
simplify and in fact may not be consistent. Note that when ${\cal{B}}=0$ and $C_2b^2+C_3=\alpha$ we regain Case 1 with ${\cal{A}}=0$ and (\ref{Aeqn}) becomes a Bernoulli equation. Then solution (\ref{solcase1}) is applicable. However, in general, when ${\cal{B}}=0$, the quadrature (\ref{quadeqn}) is applicable.

\section{Discussion}
Herrera {\et} \cite{herrera2} obtained the equation (\ref{geqn}) governing the gravitational behaviour of a radiating
spherical star undergoing shear-free gravitational collapse by imposing conformal flatness to the model.
Investigations of this model have thus far been confined to exact solutions of linearised forms of this equation.
The nonlinear behaviour of relativistic stellar models is an inherent part of realistic stars undergoing radiative
gravitational collapse. A study of the physical features of these models hinges on the solution of the governing nonlinear equations. In this paper we have presented exact solutions of the governing equation in which the nonlinearity
 has been preserved. This has been effected by transforming the equation (\ref{geqn}) into an Abel equation (\ref{calAeqn}).
  We have found several classes of exact solutions given in  (\ref{solcase1}), (\ref{solcase2a}), (\ref{solcase2b}) and (\ref{solcase3}) retaining
   the nonlinearity of the model. Note that these generate an infinite family of solutions which allow for a systematic study of radiating relativistic spheres in different scenarios.

It is important to observe that simple particular cases can be generated from our nonlinear models in Section 4. For example with ${\cal{A}}=0$ and $U\neq0$ we may obtain the line element
\be
ds^2=B^2[-dt^2+dr^2+r^2(d\theta^2+\sin^2\theta d\phi^2)]\label{linecase1}
\ee
where (\ref{linecase1}) is in conformally flat form. Here for the simple case $C_3=\alpha$ in (\ref{solcase1}) we obtain $C_1=0,\,\,C_2=0\,$ and then $B^2 $ is a constant; the Minkowski spacetime is regained. It is interesting to observe that the case $C_1=0$ in (\ref{solcase1}) also arises when $C_3$ takes the value\\
\be
C_3=\frac{2\alpha^2\beta e^{-2t/b}-\alpha}{2\alpha\beta e^{-2t/b}-3}\nonumber
\ee
where $\beta$ is a constant of integration and $C_2=(\alpha-C_3)/b^2$. Then we can write
\be
B^2=\left[\frac{b^2(2\alpha\beta e^{-2t/b}-3)}{\alpha b^2-2\alpha r^2-2b^2\alpha^2\beta e^{-2t/b}}\right]^2\nonumber
\ee
This simple analytic form facilitates the analysis of the physical features of the model. We consider now some physical features which may be investigated in
future work. With suitable choices of the arbitrary time functions,
the luminosity radius \be L=(rB)_{\Sigma}\nonumber\ee may be easily found. The quantity
\be \Gamma=\frac{d \ln p}{d \ln \rho}\nonumber\ee gives a measure of the
dynamical instability of the stellar configuration at any given
instant in time. We can use this result to confirm that the centre of the star is more unstable than the outer regions. Of particular
importance is the thermal evolution of the fluid. The causal
transport equation in the absence of rotation and viscous stress is
\be \tau
h_a^{\,b}\dot{q}_b+q_a=-\kappa(h_a^{\,b}\nabla_bT+T\dot{u}_a)
\label{transport1}\ee where $h_{ab}=g_{ab}+u_au_b$ projects into the
comoving rest space, {\it{T}} is the local equilibrium temperature, $\kappa
(\geq0)$ is the thermal conductivity, and $\tau (\geq0)$ is the
relaxation time-scale which gives rise to the causal and stable
behaviour of the theory. As shown in Maharaj and Govender
\cite{maharaj} for a physically reasonable radiative stellar model
(\ref{transport1}) becomes \be
\beta(qB)^{\dot{}}T^{-\sigma}+A(qB)=-\alpha\frac{T^{3-\sigma}(AT)^{'}}{B}\label{transport2}
\ee where $A$ and $B$ are the metric functions. Both the causal and
noncausal solutions of (\ref{transport2}) may be investigated in a simple model.

\section*{Acknowledgements}
SSM thanks the National Research Foundation and the Durban
University of Technology for financial support. SDM acknowledges
that this work is based upon research supported by the South African
Research Chair Initiative of the Department of Science and
Technology and National Research Foundation.

\end{document}